# Theoretical Limit Of Concentration Sensing of Single Receptor Artificial Biosensors

*Tuhin Chakrabortty*[\*], *Manoj M Varma*[\*#]

[\*] *Center for Nano Science and Engineering, Indian Institute of Science, Bangalore*
[#] *Robert Bosch Centre for Cyber-Physical Systems, Indian Institute of Science, Bangalore*

**Abstract**

Artificially engineered biosensors are highly inefficient in accurately measuring the concentration of biomarkers, particularly, during early diagnosis of diseases. On the other hand, single cellular systems such as chemotactic bacteria can sense their environment with extraordinary precision. Therefore, one would expect that implementing the optimal cellular sensing strategies in state-of-the-art artificial sensors can produce optimally precise biosensors. However because of the presence of measurement noise, strategies that are optimal in biological systems may not be optimal in artificial systems. Therefore, mimicking biological strategies may not be the optimal path in case of artificial sensing systems because of the presence of inherent measurement noise.

**Introduction**

Sensing their environment with high precision is crucial for single cellular organisms. Experiments reveal that some single bacteria like *E.coli* can respond to a concentration of attractants as low as 3.2nM [1]. To achieve this, cells utilize certain cell surface proteins which interacts with specific extracellular biomolecules. However, these biochemical interactions are highly stochastic in nature. How accurately can the cells measure the concentration of an extracellular molecule given the stochasticity in the binding kinetics of protein-protein interactions? This interesting question was first addressed by Berg and Purcell in their seminal work, where they derived the fundamental bounds for concentration sensing for a single cell [2]. Assuming the binding kinetics of individual receptors to be independent of each other, Berg and Purcell considered a single receptor sensor and argued that the results could be extended for multiple receptors. They described a time averaging-based sensing strategy, where the cell estimates the concentration of the ligands by calculating the average time the receptor spends being bound. For such a system, the average must be calculated from the binary time series data of binding-unbinding events of the receptor over a finite integration time interval ($T$). Berg and Purcell claimed that the estimation error due to this finiteness of $T$ defines the theoretical lower limit for the concentration sensing by a single receptor sensing system which they showed to be

$$\frac{\Delta c_{rms}}{\bar{c}} = \left(\frac{2\tau_b}{T\bar{p}}\right)^{1/2} \qquad (1)$$

Where $\Delta c_{rms}$ is the root mean square error in estimation, $\bar{c}$ is the concentration to be estimated, $p$ represents the occupancy of the receptor which is 1 when the receptor is bound and 0 when it is unbound. $\bar{p}$ is the time average occupation. $\tau_b$ is the average time, a ligand molecule stays bound to a receptor.

Similar to cellular sensing systems, artificially engineered biosensors mostly rely on interactions between receptors and a set of specific biomarkers secreted by the infected cells. However, unlike cells, these artificial biosensors are highly inefficient, particularly during early stages of the diseases [3]. Low concentration of the biomarkers during the early stage, makes the sensing extremely challenging. One way to address this challenge would be to mimic cellular sensing strategies in the state-of-the-art artificial sensors. The limit of sensing of such an artificial system at equilibrium should be expected to reach the limit of single cellular sensing described by Berg and Purcell and subsequent works [2, 4, 5] However, besides the stochasticity due to binding kinetics of receptor-ligand interaction as in case of biological systems, there are other unavoidable sources of stochasticity in artificial sensing systems. For example, the measurement techniques used in these systems are inherently noisy. Therefore, the assumption of a binary output due to the binding and unbinding events in case of a single receptor sensing system is not valid for artificial sensors. Instead the output of these sensors are noisy versions of binary signals.

With this constraint, what is the fundamental limit of concentration sensing for these artificial biosensors? More importantly, are the optimal sensing strategies for cellular sensing systems still remain optimal for artificial sensing systems? To address these questions, we have derived the theoretical limit of sensing of an artificial biosensor with the Berg and Purcell strategy ($BP$). We have also designed a novel autocorrelation-based signal processing strategy for single receptor artificial sensors. We demonstrate that compared to $BP$ our strategy performs better for estimating concentration in the presence of measurement noise.

**Limit of sensing for artificial biosensors**

1. **Berg and Purcell strategy**

We consider an artificially engineered sensor with a single receptor. Contrary to the binary signal approximation in previous works, we define the signal $(m(t))$ observed by the sensor as

$$m(t) = \mathrm{B}(t) + \xi(t) \qquad (2)$$

Where $\mathrm{B}(t)$ is the binary signal with transition probabilities for $1 \to 0$ and $0 \to 1$ proportional to $k_+ c$ and $k_-$ respectively, where $c$ is the concentration of the ligand molecules and $k_+$ and $k_-$ are the binding and unbinding rates of the ligand molecules with the receptor respectively. The noise term $(\xi(t))$ is independent of the binary signal $\mathrm{B}(t)$. Although $\xi(t)$ is ideally assumed to be zero mean Gaussian white noise, the arguments presented here are valid for any zero mean continuous probability distribution with a finite persistence time $(\tau_\xi)$, which can be defined as $\langle \xi(t) \rangle = 0$ and $\langle \xi(t')\xi(t'+\tau) \rangle|_{\tau > \tau_\xi} = 0$.

Using Berg and Purcell strategy, the time average of the signal $m(t)$ can be calculated as

$$m_T = \frac{1}{T}\int_a^{a+T} m(t)dt = \overline{B} = \frac{c}{c + K_D} \tag{3}$$

where $K_D = k_-/k_+$, $\overline{B} = \langle B(t) \rangle$ and T is the estimation time

Due to finiteness of estimation time (*T*), the estimation error can be calculated by calculating the variance in $m_T$ for different $a$ but same T.

$$\langle m_T^2 \rangle = \frac{1}{T^2} \int_0^T dt' \int_0^T dt\, G(t - t')$$

where $G(\tau)$ is the autocorrelation of the signal $m(t)$ defined as

$$G(\tau) = \langle m(t)m(t+\tau) \rangle$$

$$= \langle B(t)B(t+\tau) \rangle + \langle \xi(t)\xi(t+\tau) \rangle$$

If $T \gg \tau_b$, where $\tau_b$ is the average time interval for which a molecule stays bound to the receptor, then, following the derivation in Berg and Purcell,

$$\langle m_T^2 \rangle - \langle m_T \rangle^2 = \frac{2}{T}\overline{B}(1-\overline{B})^2 \tau_b + \frac{1}{T^2}\int_0^T dt' \int_0^T dt\, \langle \xi(t)\xi(t') \rangle \tag{4}$$

The first term of eq (4) is same as the eq (50) in Berg and Purcell. The second term is due to the noise added to the system. If $\xi(t)$ is a white noise, then eq (4) becomes

$$\Delta m_T^2 = \frac{2}{T}\overline{B}(1-\overline{B})^2 \tau_b + \frac{\sigma_\xi^2}{T} \tag{5}$$

Equation (5) represents the error in estimating $m_T$. The concentration $c$ is calculated from $m_T$ using equation (3). Because of propagation of error, the error in estimation of $c$ becomes

$$\left(\frac{\Delta c}{c}\right)_{BP} = \sqrt{\frac{1}{Tm_T}\left(2\tau_b + \frac{\sigma_\xi^2}{m_T(1-m_T)^2}\right)} \tag{6}$$

**Autocorrelation based sensing strategy**

We now present an autocorrelation-based sensing strategy that is independent of the measurement noise. In this signal processing technique, we calculate the autocorrelation of the signal $m(t)$ with a shifted version $m(t + \tau_s)$ instead of the time average in case of Berg and Purcell. The autocorrelation $G_T(\tau_s)$ of the signal $m(t)$ over a finite time period $T$ can be written as

$$G_T(\tau_s) = \frac{1}{T}\int_a^{a+T} m(t)m(t+\tau_s)dt$$

If $\tau_s > \tau_\xi$, $\langle \xi(t)\xi(t+\tau_s)\rangle = 0$. Therefore, $G_T(\tau_s)$ is independent of the measurement noise. The expression of $G_T(\tau_s)$ becomes

$$G_T(\tau_s) = \frac{1}{T}\int_a^{a+T} B(t)B(t+\tau_s)dt$$

From the derivation in Berg and Purcell the estimated autocorrelation $G_T(\tau_s)$ will become [2],

$$G_T(\tau_s) = \overline{B}^2 + \overline{B}(1-\overline{B})\exp(-|\tau_s|/(1-\overline{B})\tau_b)$$

where $\tau_b$ is the average bound time and can be defined as $\tau_b = 1/k_-$

However, similar to Berg and Purcell, due to finiteness of estimation time ($T$), there will be estimation error in $G_T$, which can be calculated as

$$\langle G_T^2\rangle = \frac{1}{T^2}\int_0^T dt' \int_0^T dt\, \Gamma(t-t')$$

where $\Gamma(\tau) = \langle B(t)B(t+\tau_s)B(t+\tau)B(t+\tau+\tau_s)\rangle$

for $T \gg \tau_b$, the estimation error $\Delta G_T^2(\tau_s)$ becomes

$$\Delta G_T^2(\tau_s) = \langle G_T^2\rangle - \langle G_T\rangle^2 = \left(\frac{2\tau_b}{T}\right)G_T(\tau_s)[1-G_T(\tau_s)](1-\overline{B}) \quad (7)$$

Unlike the Berg and Purcell strategy, the estimation error of the autocorrelation $G_T$ doesn't depend only depends on the persistence time of the noise signal and not on the variance. Therefore, for a white measurement noise, as assumed in equation (5), the estimation error can be calculated by taking $\tau_s \ll \tau_b$ as

$$\left(\frac{\Delta c}{c}\right)_{autocorr} = \sqrt{\left(\frac{1-G_T(\tau_s)}{1-m_T}\right)\left(\frac{2\tau_b}{TG_T(\tau_s)}\right)} \quad (8)$$

For $\tau_s = 0$, we can retrieve the result derived by Berg and Purcell.

**Comparison of Autocorrelation-based strategy with Berg and Purcell strategy:**

For a low measurement noise conditions such as cellular sensing, the Berg and Purcell strategy performs better than the autocorrelation-based strategy. However, the estimation error in the Berg and Purcell strategy is highly dependent on the variance of the measurement noise. Therefore, with increase in measurement noise, the estimation error exponentially increases [Fig 1]. On the other hand, the estimation error in case of autocorrelation-based strategy is independent of the measurement noise. Therefore, for high measurement noise scenarios as seen in case of artificial sensing systems, autocorrelation-based strategy outperforms the Berg and Purcell strategy.

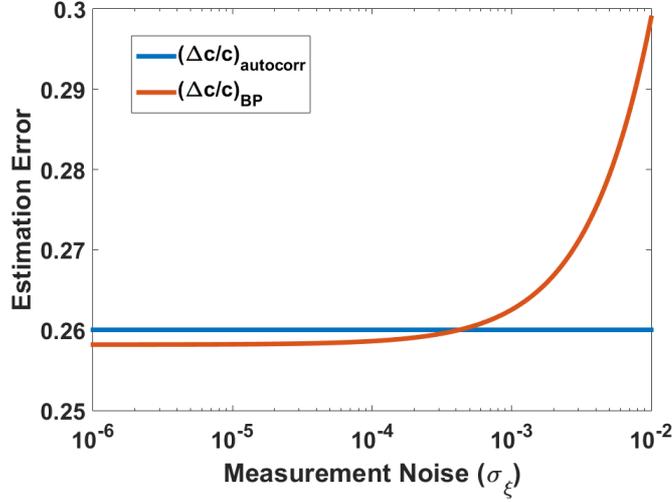

Figure 1: Effect of measurement noise on the estimation error for BP and autocorrelation-based strategies. For low measurement noise, BP performs better. However, as noise increases, the estimation error in BP exponentially increases. Therefore, for highly noisy systems, autocorrelation based method is a better strategy than BP. [Parameters Used: $T = 10, \tau_b = 0.1, \tau_s = 0.001, m_T = 0.3$]

**Discussion**

In this work, we have calculated the limit of concentration sensing of a model artificial biosensor with a single receptor. The model system mimics the working principle of single nanopore sensor and hence has practical relevance. In case of single nanopore sensors, molecules are passed through a nanopore and the dip in the current across the pore is measured [6, 7]. By calculating the number of the dips in the current signal, one can determine the concentration of the molecules in the solution. Given the temporal information of the current across the pore, what is the optimal strategy to extract the information of the concentration of the molecules in the solution? One method would be to use the Berg and Purcell strategy by calculating the time average of the signal and mapping the results to the concentration of the molecules. However, due to the dimensional constraints the measurement system in this case is inherently noisy. Therefore, the estimation error will be very high as described in Figure 1. On the other hand, if the autocorrelation of the temporal signal is calculated instead of the time average then the estimation error becomes independent of the measurement noise and therefore will provide a better strategy than the Berg and Purcell strategy in the presence of high measurement noise. This work demonstrates that mimicking biological strategies may not be the optimal path in case of artificial sensing systems because of the presence of inherent measurement noise.

**References:**

[1] R. M. Macnab and D. Koshland, "The gradient-sensing mechanism in bacterial chemotaxis," *Proceedings of the National Academy of Sciences*, vol. 69, no. 9, pp. 2509–2512, 1972.

[2] H. C. Berg and E. M. Purcell, "Physics of chemoreception," *Biophysical journal*, vol. 20, no. 2, pp. 193–219, 1977.

[3] S. S. Hori and S. S. Gambhir, "Mathematical model identifies blood biomarker-based early cancer detection strategies and limitations," vol. 3, pp. 109ra116–109ra116, 2011.

[4] P. R. ten Wolde, N. B. Becker, T. E. Ouldridge, and A. Mugler, "Fundamental limits to cellular sensing," vol. 162, pp. 1395–1424, 2016.

[5] K. Kaizu, W. de Ronde, J. Paijmans, K. Takahashi, F. Tostevin, and P. ten Wolde, "The berg-purcell limit revisited," vol. 106, pp. 976–985, 2014.

[6] Y. Wang, D. Zheng, Q. Tan, M. Wang, and L.-Q. Gu, "A nanopore sensor for single molecule detection of circulating micrornas in lung cancer patients," vol. 100, p. 168a, 2011.

[7] K. Tian and L.-Q. Gu, "Nanopore single-molecule dielectrophoretic detection of cancer-derived microrna biomarkers," 2013.